\documentclass[preprint,amsmath,amssymb,aps,showkeys,showpacs]{revtex4}
\usepackage[english]{babel}
\usepackage{graphicx}
\usepackage{graphics}
\usepackage{amsmath}
\usepackage{dcolumn}
\usepackage{amssymb}
\usepackage{bm}

\begin{document}
\title{Relativistic scalar particle subject to a confining potential and Lorentz symmetry breaking effects in the cosmic string spacetime}
\author{H. Belich} 
\email{belichjr@gmail.com}
\affiliation{Departamento de F\'isica e Qu\'imica, Universidade Federal do Esp\'irito Santo, Av. Fernando Ferrari, 514, Goiabeiras, 29060-900, Vit\'oria, ES, Brazil.}

\author{K. Bakke}
\email{kbakke@fisica.ufpb.br}
\affiliation{Departamento de F\'isica, Universidade Federal da Para\'iba, Caixa Postal 5008, 58051-970, Jo\~ao Pessoa, PB, Brazil.}

\begin{abstract}
The behaviour of a relativistic scalar particle subject to a scalar potential under the effects of the violation of the Lorentz symmetry in the cosmic string spacetime is discussed. It is considered two possible scenarios of the Lorentz symmetry breaking in the CPT-even gauge sector of the Standard Model Extension defined by a tensor $\left(K_{F}\right)_{\mu\nu\alpha\beta}$. Then, by introducing a scalar potential as a modification of the mass term of the Klein-Gordon equation, it is shown that the Klein-Gordon equation in the cosmic string spacetime is modified by the effects of the Lorentz symmetry violation backgrounds and bound state solution to the Klein-Gordon equation can be obtained.

\end{abstract}

\maketitle

\section{Introduction}

The Standard Model (SM) is the basis of our understanding of the physics of fundamental interactions, despite the gravitational interaction is not included. The recent discovery of the Higgs boson at the LHC endorse the research program which aims to explain the physics of fundamental interactions as manifestations of excitations of fundamental fields, therefore, the SM can also explains the origin of mass of the particles. Although this tremendous progress in the description of particle physics, it is known that the description of massless neutrinos presents in the Standard Model is not satisfactory. Furthermore, the origin of electron electric dipole moment has not been explained by Standard Model, where just experimental upper bounds have been established \cite{revmod}. By the Standard Model, an upper limit for electron electric dipole moment was established as $d_{e}\leq 10^{-38}\,\mathrm{e}\cdot\mathrm{cm}$ \cite{revmod}. However, experimental measurements yielded an upper limit given by $d_{e}\leq10^{-29}\,\mathrm{e}\cdot\mathrm{cm}$ by using a polar molecule thorium monoxide (ThO) \cite{science}. This experimental result has shown us a necessity of investigating the physics beyond the Standard Model since the term associated with the electric dipole moment violates the CP symmetry. 

In fact, the SM is a effective theory, where it is expected that a new physics may appear when the LHC reach $14\mathrm{TeV}$. At this point, we might question what concepts that could be used as guides for a more fundamental theory. By taking as example the electroweak unification by the SM, the Higgs mechanism is an interesting way to get clues about how to obtain the fundamental theory. A proposed scenario beyond the Standard Model is the extension of the mechanism for spontaneous symmetry breaking through vector or tensor fields. It implies that the Lorentz symmetry is violated. This proposal has been established after the seminal work made by Kosteleck\'{y} and Samuel \cite{extra3} in a string field theory, where it is shown that the Lorentz symmetry is violated through a spontaneous symmetry breaking mechanism triggered by the appearance of nonvanishing vacuum expectation values of nontrivial Lorentz tensors. A general framework for testing the low-energy manifestations of CPT symmetry and the Lorentz symmetry breaking is known as the Standard Model Extension (SME) \cite{colladay-kost}. In this framework, the effective Lagrangian corresponds to the usual Lagrangian of the Standard Model to which is added to the Standard Model operators a Lorentz violation tensor background. In this way, The effective Lagrangian is written as being an invariant under the Lorentz transformation of coordinates in order to guarantee that the observer independence of physics. However, the physically relevant transformations are those that affect only the dynamic fields of the theory \cite{coll-kost,baeta,bras}. Empirical studies in the flat spacetime background include muons \cite{muon}, mesons \cite{meson,meson2}, baryons \cite{barion}, photons \cite{photon,kost2}, electrons \cite{electron}, neutrinos\cite{neutrino} and the Higgs \cite{higgs} sector. The gravity sector has also been explored in Refs. \cite{curv,gravity,gravity2,curv2,curv3,bb15}. In Ref. \cite{data}, one can find the current limits on the coefficients of the Lorentz symmetry violation. It is worth mentioning other studies of Lorentz symmetry breaking effects, for instance, in Weyl semi-metals \cite{weyl}, in tensor backgrounds \cite{louzada,manoel2}, Rashba coupling \cite{rash,bb3}, in a quantum ring \cite{bb5}, in the quantum Hall effect \cite{lin2} and geometric quantum phases \cite{belich,bb2,bb4,lbb}.

In this paper, we investigate the behaviour of a relativistic scalar particle subject to a scalar potential under the effects of the violation of the Lorentz symmetry in the cosmic string spacetime. We study two possible scenarios of the anisotropy generated by a Lorentz symmetry breaking term defined by a tensor $\left(K_{F}\right)_{\mu\nu\alpha\beta}$ that corresponds to a tensor that governs the Lorentz symmetry violation in the CPT-even gauge sector of the Standard Model Extension. We introduce the scalar potential by modifying the mass term of the Klein-Gordon equation, and show that the Klein-Gordon equation in the cosmic string spacetime is modified by the effects of the Lorentz symmetry violation backgrounds and bound state solution to the Klein-Gordon equation can be obtained.

This paper is organized as follows: in section II, we make a brief introduction to the modified Maxwell theory coupled to gravity that allows us to obtain an effective metric for a curved spacetime under Lorentz symmetry breaking effects; in section III, we introduce a scalar potential by modifying the mass term of the Klein-Gordon equation and define a fixed space-like 4-vector that yields an effective metric of the cosmic string spacetime under Lorentz symmetry breaking effects; thus, we obtain the bound state solutions to the Klein-Gordon equation and the allowed energies of the relativistic system; in section IV, we define a fixed time-like 4-vector and, by introducing the scalar potential as a modification of the mass term, then, we obtain the allowed energies of the system; in section V, we present our conclusions.

\section{effective metric yield by Lorentz symmetry breaking effects}

In this section, we make a brief introduction to the modified Maxwell theory coupled to gravity that allows us to obtain an effective metric for the cosmic string spacetime under Lorentz symmetry breaking effects. Investigations in topics such as the violation of the CPT-symmetry and birefringence in vacuum have been examples of the possibilities of building the extension of the Standard Model (SME). The CPT-even gauge sector of the SME has been studied since 2002, after the pioneering contributions made by Kosteleck\'y and Mewes \cite{19,20}, and there is an extensive literature dealing with the extension of the Standard Model in the even sector of SME by the following term \cite{cas}:
\begin{equation}
S=-\frac{1}{4}\int d^{4}x\;K_{abcd}\,F^{ab}\,F^{cd}. 
\label{1.1}
\end{equation}

The tensor $K_{abcd}$ given in Eq. (\ref{1.1}) does not violate the CPT-symmetry. Despite the violation of the CPT-symmetry implies that the Lorentz invariance is violated \cite{greenberg}, the reverse is not necessarily true, because the action given in Eq. (\ref{1.1}) breaks the Lorentz symmetry in the sense that the tensor $K_{abcd}$ has a non-null vacuum expectation value. It is worth mentioning that tensor $K_{abcd}$ has the same properties of the Riemann tensor, as well as an additional double-traceless condition. This tensor also possesses the following symmetries: $K_{abcd}=K_{\left[ab\right]\left[cd\right]}$, $K_{abcd}=K_{cdab}$ and $K^{ab}_{\,\,\,\,\,\,\,ab}=0$. In addition, by following Refs. \cite{curv3,curv2,curv4}, we can write the tensor $K_{abcd}$ in terms of a traceless and symmetric matrix $\tilde{\kappa}_{ab}$ as
\begin{equation}
K_{abcd}=\frac{1}{2}\left(\eta_{ac}\,\tilde{\kappa}_{bd}-\eta _{ad}\,\tilde{\kappa}_{bc}+\eta_{bd}\,\tilde{\kappa}_{ac}-\eta_{bc}\tilde{\kappa}_{ad}\right).
\label{1.3}
\end{equation}

An interesting point of discussion is that by defining a normalized parameter 4-vector $\xi^{a}$, which must satisfy the conditions $\xi_{a}\xi^{a}=1$ for the timelike case and $\xi_{a}\xi^{a}=-1$ for the spacelike case, thus, we can decompose the tensor $\tilde{\kappa}_{ab}$ in the following form:
\begin{equation}
\tilde{\kappa}_{ab}=\kappa\left(\xi_{a}\xi_{b}-\frac{\eta_{ab}\,\xi ^{c}\xi_{c}}{4}\right),
\label{1.4}
\end{equation}
where $\kappa=\frac{4}{3}\tilde{\kappa}^{ab}\,\xi_{a}\,\xi_{b}$. As in Refs. \cite{curv3,curv4}, we consider the parameter $\kappa$ to be a spacetime independent parameter, where $0\leq\kappa<2$.

Recently, it has been pointed out in Refs. \cite{curv2,curv3} that the event horizon of a black hole is modified by the anisotropy generated by $K_{\mu\nu\kappa\lambda}$. Therefore, an interesting way of analysing new phenomenology for verification Lorentz symmetry violation is to consider a curved spacetime background \cite{curv}. From this perspective, let us write the corresponding Lagrange density to the nonbirefringent modified Maxwell theory coupled to gravity \cite{curv2,curv3}:
\begin{equation}
\mathcal{L}_{\mathrm{modM}}=-\sqrt{g}\,\left(\frac{1}{4}\,F_{\mu\nu}F_{\rho\sigma}\,g^{\mu\rho}g^{\nu\sigma}+\frac{1}{4}\,K^{\mu\nu\rho\lambda}\,F_{\mu\nu}F_{\rho\lambda}\right).
\label{1.5}
\end{equation}

Thereby, by using the relations mentioned above, the Lagrange density (\ref{1.5}) can be written in terms of an effective metric tensor $\bar{g}_{\mu\nu}\left(x\right)$ as
\begin{eqnarray}
\mathcal{L}_{\mathrm{modM}}=-\sqrt{g}\left(1-\frac{1}{2}\,\kappa\,\xi_{\alpha}\,\xi^{\alpha}\right)\,\frac{1}{4}\,F^{\mu\nu}\left(x\right)F^{\rho\sigma}\left(x\right)\,\bar{g}_{\mu\rho}\left(x\right)\,\bar{g}_{\nu\sigma}\left(x\right),
\label{1.6}
\end{eqnarray}
where the expression of this effective metric tensor is given by \cite{curv2,curv3}:
\begin{eqnarray}
\bar{g}_ {\mu\rho}\left(x\right)=g_{\mu\rho}\left(x\right)-\epsilon\,\xi_{\mu}\,\xi_{\rho},
\label{1.7}
\end{eqnarray}
whose parameter $\epsilon$ is defined as $\epsilon=\frac{\kappa}{1+\frac{\kappa}{2}}$ and $\bar{g}^{\mu\nu}\,\bar{g}_{\nu\alpha}=\delta^{\mu}_{\,\,\,\alpha}$. However, all lowering or raising of indices is performed by using the original background metric $g_{\mu\nu}\left(x\right)$ and its inverse $g^{\mu\nu}\left(x\right)$. This type of background yields the anisotropy in the spacetime \cite{curv}, hence, this suggests that the propagation of the photon must be modified by this background.

In recent years, a geometrical approach to study Lorentz symmetry breaking effects has been proposed based on the Kaluza-Klein theory, by considering the Lorentz violating tensor fields with expectation values along the extra directions \cite{ct,petrov}. Based on this perspective, our proposal is to extend the geometric approximation that gives rise to the effective metric tensor (\ref{1.7}) to a fermionic field by modifying the cosmic string spacetime. In Refs. \cite{bh,bh2,bh3,curv2,curv3} is discussed that the generalized second law of thermodynamics would be violated by the modified Maxwell theory and fermions in the presence of a black hole. However, in the cosmic string spacetime there is no event horizons, then, if the cosmic string spacetime is modified by the effects of a Lorentz symmetry breaking background, therefore there is nothing, at first principles, that prohibit fields and particles described by the Standard Model to feel the anisotropies described by the effective metric tensor (\ref{1.4}). Our proposal is to analyze the spontaneous violation of Lorentz symmetry as an effective theory that can suggest how to move in the direction of a fundamental theory. Therefore, we do not analyze limit situations where singularities appear, such as the event horizon, because in fact we are not sure that these are real situations in a more fundamental theory. By assuming that this extension is possible, the aim of this work is to investigate quantum effects on a relativistic scalar particle subject to a scalar potential yielded by the effects of the Lorentz symmetry violation in the cosmic string spacetime. In the next sections, we discuss two particular cases that allows to obtain bound states solutions to the Klein-Gordon equation in the cosmic string spacetime under Lorentz symmetry breaking effects.

\section{fixed space-like 4-vector case}

In this work, we wish to study relativistic quantum effects that stem from a Lorentz symmetry breaking effects in a topological defect spacetime by using the geometrical picture given by the effective metric tensor (\ref{1.7}). For this purpose, let us consider the cosmic string spacetime, whose line element is given by
\begin{eqnarray}
ds^{2}=dt^{2}-d\rho^{2}-\eta^{2}\rho^{2}d\varphi^{2}-dz^{2},
\label{1.8}
\end{eqnarray}
where the parameter $\eta$ is related to the deficit angle and it is defined as $\eta=1-4\varpi G/c^{2}$, with $\varpi$ being the linear mass density of the cosmic string. The azimuthal angle varies in the interval $0\leq\varphi<2\pi$. The deficit angle can assume only values in which $0<\eta<1$, because values greater than 1 correspond to an anti-cone with negative curvature, which makes sense only in the description of linear defects in solid crystal \cite{kat,furt}. It is worth mentioning that linear topological defects can be viewed in the solid state context as the analogue of three-dimensional gravity through the formalism proposed by Katanaev and Volovich \cite{kat}. On the other hand, the appearance of topological defects in the spacetime, such as the cosmic string (\ref{1.10}), is considered to be through phase transitions during the evolution of the Universe which involves a symmetry breaking \cite{td2}. In general, a defect corresponds to singular curvature, torsion or both along the line defect. It must be emphasized that this geometry has a curvature tensor which represents a conical singularity $R_{\rho,\varphi}^{\rho,\varphi}=\frac{1-\eta}{4\eta}\,\delta_{2}(\vec{r})$, where $\delta_{2}(\vec{r})$ is the two-dimensional delta function. This behavior of the curvature tensor is denominated conical singularity \cite{staro} because it gives rise to the curvature concentrated on the cosmic string axis, with all other places having null curvature.

In this section, let us consider the normalized parameter four-vector $\xi_{a}$ as a space-like 4-vector given by \cite{bb15}
\begin{eqnarray}
\xi_{a}=\left(0,0,1,0\right).
\label{1.9}
\end{eqnarray}

In this way, we can write the four-vector $\xi_{\mu}\left(x\right)$ given in the effective metric tensor (\ref{1.7}) via relation: $\xi_{\mu}\left(x\right)=e^{a}_{\,\,\,\mu}\left(x\right)\,\xi_{a}=\eta\rho\,\xi_{2}$; thus, we have that the condition $\xi_{\mu}\left(x\right)\,\xi^{\mu}\left(x\right)=\mathrm{const}$ established in Refs. \cite{curv,curv2,curv3} is satisfied, and the effective metric of the cosmic string spacetime under Lorentz symmetry breaking effects becomes
\begin{eqnarray}
\bar{ds}^{2}=dt^{2}-d\rho^{2}-\eta^{2}\rho^{2}\left(1+\epsilon\right)d\varphi^{2}-dz^{2}.
\label{1.10}
\end{eqnarray}

Recently, a great deal of works has investigated the behaviour of a relativistic scalar particle subject to scalar confining potentials \cite{kg,kg2,kg3,kg4,greiner,scalar,scalar2}. A interesting point raised in these studies is the way of introducing a scalar potential into the Klein-Gordon equation. By following Ref. \cite{greiner}, a scalar potential can be introduced into the Klein-Gordon equation by modifying the momentum operator $p_{\mu}=i\partial_{\mu}$ via relation $p_{\mu}\rightarrow p_{\mu}-q\,A_{\mu}\left(x\right)$, that is, in the same way of introducing the electromagnetic 4-vector potential into the Klein-Gordon equation. On the other hand, in Ref. \cite{scalar} is shown that a scalar potential can be introduced into the Klein-Gordon equation by making a modification in the mass term as $m\rightarrow m+S\left(\vec{r},\,t\right)$, where $S\left(\vec{r},\,t\right)$ is the scalar potential. It is worth mentioning some works that have explored this modification of the mass term such as the scalar field in the cosmic string spacetime \cite{eug}, a Dirac particle in the presence of static scalar potential and a Coulomb potential \cite{scalar2} and the quark-antiquark interaction as a problem of a relativistic spin-$0$ possessing a position-dependent mass \cite{bah}. Other works that have explored the relativistic quantum dynamics of a scalar particle subject to different confining potentials which can be in the interest of several areas of physics \cite{alvaro,qian,castro,alhardi,adame,xu}. By following Ref. \cite{quark}, the introduction of a Coulomb-like potential as an additional term of the mass of the particle has a particular interest in studies of confinement of quarks, such as in the {\it bag} model, since it can generalize the rest mass of the particle to the model of confinement of quarks. Therefore, let us consider a scalar potential given by
\begin{eqnarray}
S\left(\rho\right)=\frac{\chi}{\rho},
\label{1.11}
\end{eqnarray}
where $\chi$ is a parameter that characterizes the scalar potential.

In this work, we introduce the scalar potential by modifying the mass term of the Klein-Gordon equation, then, the Klein-Gordon equation in a curved spacetime background is written as \cite{bd,eug}:
\begin{eqnarray}
\left[m+S\left(\rho\right)\right]^{2}\phi=\frac{1}{\sqrt{-g}}\,\partial_{\mu}\left[\sqrt{-g}\,g^{\mu\nu}\,\partial_{\nu}\right]\phi,
\label{1.12}
\end{eqnarray}
where $g=\mathrm{det}\left(g_{\mu\nu}\right)$. As discussed in the previous section, we can write the Klein-Gordon equation (\ref{1.12}) in terms of the effective metric tensor $\bar{g}_{\mu\rho}\left(x\right)$ by using the line element given in Eq. (\ref{1.10}) and thus analyse the effects of the Lorentz symmetry violation on a relativistic scalar particle subject to the scalar potential (\ref{1.11}) from a geometrical point of view. Thereby, by changing $g_{\mu\nu}\left(x\right)\rightarrow\bar{g}_{\mu\nu}\left(x\right)$, the Klein-Gordon equation (\ref{1.12}) becomes
\begin{eqnarray}
\left[m+\frac{\chi}{\rho}\right]^{2}\phi=-\frac{\partial^{2}\phi}{\partial t^{2}}+\frac{\partial^{2}\phi}{\partial\rho^{2}}+\frac{1}{\rho}\frac{\partial\phi}{\partial\rho}+\frac{1}{\rho^{2}\eta^{2}\left(1+\epsilon\right)}\frac{\partial^{2}\phi}{\partial\varphi^{2}}+\frac{\partial^{2}\phi}{\partial z^{2}}
\label{1.13}
\end{eqnarray}

Henceforth, let us consider a particular solution to Eq. (\ref{1.13}) given by the eigenfunctions of the operators $\hat{p}_{z}=-i\partial_{z}$ and $\hat{L}_{z}=-i\partial_{\varphi}$. Therefore, we can write the solution to Eq. (\ref{1.13}) in terms of the eigenvalues of these operators as follows:
\begin{eqnarray}
\phi=e^{-i\mathcal{E}t}\,e^{il\varphi}\,e^{ikz}\,f\left(\rho\right),
\label{1.14}
\end{eqnarray}
 where $l=0,\pm1,\pm2,\ldots$, $k$ is a constant and $f\left(\rho\right)$ is a function of the radial coordinate. From now on, we simplify our system by considering $k=0$. Then, substituting (\ref{1.14}) into Eq. (\ref{1.13}), we obtain
\begin{eqnarray}
\frac{d^{2}f}{d\rho^{2}}+\frac{1}{\rho}\frac{df}{d\rho}-\frac{\gamma^{2}}{\rho^{2}}\,f-\frac{\delta}{\rho}\,f-\beta^{2}\,f=0,
\label{1.15}
\end{eqnarray}
where we have defined the following parameters:
\begin{eqnarray}
\beta^{2}&=&m^{2}-\mathcal{E}^{2};\nonumber\\
\delta&=&2m\chi;\label{1.16}\\
\gamma^{2}&=&\frac{l^{2}}{\eta^{2}\left(1+\epsilon\right)}+\chi^{2}\nonumber.
\end{eqnarray}

Note that the fourth term of Eq. (\ref{1.16}) plays the role of an attractive Coulomb-type potential if we consider $\chi>0$. Now, let us perform a change of variables given by $\zeta=2\beta\rho$. Then, we have
\begin{eqnarray}
\frac{d^{2}f}{d\zeta^{2}}+\frac{1}{\zeta}\frac{df}{d\zeta}-\frac{\gamma^{2}}{\zeta^{2}}\,f-\frac{\delta}{2\tau\zeta}\,f-\frac{1}{4}\,f=0.
\label{1.17}
\end{eqnarray}

By analysing the asymptotic behaviour of Eq. (\ref{1.17}), we can write the solution to Eq. (\ref{1.17}) in terms of a unknown function $F\left(\zeta\right)$ as: 
\begin{eqnarray}
f\left(\xi\right)=e^{-\frac{\zeta}{2}}\,\zeta^{\left|\gamma\right|}\,F\left(\zeta\right).
\label{1.18}
\end{eqnarray}
Thereby, substituting Eq. (\ref{1.18}) into Eq. (\ref{1.17}) we obtain the following second-order differential equation:
\begin{eqnarray}
\zeta\,\frac{d^{2}F}{d\zeta^{2}}+\left[2\left|\gamma\right|+1-\zeta\right]\frac{dF}{d\zeta}+\left[\frac{\delta}{2\beta}-\left|\gamma\right|-\frac{1}{2}\right]F=0,
\label{1.19}
\end{eqnarray}
which is called as the confluent hypergeometric equation \cite{abra,arf}; thus, the function $F\left(\zeta\right)$ is the confluent hypergeometric function, that is, $F\left(\zeta\right)=\,_{1}F_{1}\left(\left|\gamma\right|+\frac{1}{2}-\frac{\delta}{2\beta},\,2\left|\gamma\right|+1,\,\zeta\right)$. It is well-known that the confluent hypergeometric series becomes a polynomial when $\left|\gamma\right|+\frac{1}{2}-\frac{\delta}{2\beta}=-n$, where $n=0,1,2,\ldots$. In this way, the corresponding relativistic energy levels of the bound states solutions are given by
\begin{eqnarray}
\mathcal{E}_{n,\,l}=\pm\,\sqrt{m^{2}-\frac{4m^{2}\chi^{2}}{\left[2n+2\left|\frac{l^{2}}{\eta^{2}\left(1+\epsilon\right)}+\chi^{2}\right|+1\right]^{2}}}.
\label{1.20}
\end{eqnarray}

Hence, Eq. (\ref{1.20}) are the allowed energy of a relativistic scalar particle confined to the scalar potential given in Eq. (\ref{1.11}) subject to the effects of the Lorentz symmetry violation in the cosmic string spacetime. In the present case, we have established a particular background of the violation of the Lorentz symmetry defined by a fixed space-like 4-vector given in Eq. (\ref{1.9}) in which allows us to write an effective metric for the cosmic string spacetime. Observe that the effects of the Lorentz symmetry violation in the relativistic energy levels (\ref{1.20}) is given by the presence of the parameter $\epsilon$, and the effects of the topology of the cosmic string spacetime is given by the presence of the parameter $\eta$. By taking the limit as $\eta\rightarrow1$ in Eq. (\ref{1.20}), then, we have the allowed energies of the relativistic scalar particle confined to the the scalar potential (\ref{1.11}) under Lorentz symmetry breaking effects in the Minkowski spacetime. On the other hand, by taking $\epsilon=0$, the effects of the Lorentz symmetry violation vanish and the relativistic energy levels become the allowed energies of the relativistic scalar particle confined to the the scalar potential (\ref{1.11}) in the cosmic string spacetime \cite{eug}. Moreover, note that the allowed energies (\ref{1.20}) do not change by performing the change $\chi\rightarrow-\chi$.

\section{fixed time-like 4-vector case}

In this section, we investigated the Lorentz symmetry breaking effects on the relativistic scalar particle yielded by the fixed time-like 4-vector $\xi_{a}$ given by
\begin{eqnarray}
\xi_{a}=\left(1,0,0,0\right).
\label{2.1}
\end{eqnarray}

Again, the condition $\xi_{\mu}\left(x\right)\,\xi^{\mu}\left(x\right)=\mathrm{const}$, established in Refs. \cite{curv,curv2,curv3} is satisfied, and the effective metric of the cosmic string spacetime under Lorentz symmetry breaking effects becomes
\begin{eqnarray}
\bar{ds}^{2}=\left(1-\epsilon\right)dt^{2}-d\rho^{2}-\eta^{2}\rho^{2}\,d\varphi^{2}-dz^{2}.
\label{2.2}
\end{eqnarray}

By considering the confinement of the relativistic scalar particle to the scalar potential given in Eq. (\ref{1.11}), the Klein-Gordon equation (\ref{1.12}) can be written as
\begin{eqnarray}
\left[m+\frac{\chi}{\rho}\right]^{2}\phi=-\frac{1}{\left(1-\epsilon\right)}\frac{\partial^{2}\phi}{\partial t^{2}}+\frac{\partial^{2}\phi}{\partial\rho^{2}}+\frac{1}{\rho}\frac{\partial\phi}{\partial\rho}+\frac{1}{\rho^{2}\eta^{2}}\frac{\partial^{2}\phi}{\partial\varphi^{2}}+\frac{\partial^{2}\phi}{\partial z^{2}}
\label{2.3}
\end{eqnarray}

A particular solution to Eq. (\ref{2.3}) can also be given as in Eq. (\ref{1.14}), that is, $\phi=e^{-i\mathcal{E}t}\,e^{il\varphi}\,e^{ikz}\,h\left(\rho\right)$, where $h\left(\rho\right)$ is a function of the radial coordinate; thus, the Klein-Gordon equation (\ref{2.3}) becomes
\begin{eqnarray}
\frac{d^{2}h}{d\rho^{2}}+\frac{1}{\rho}\frac{dh}{d\rho}-\frac{\tau^{2}}{\rho^{2}}\,h-\frac{\delta}{\rho}\,h-\mu^{2}\,h=0,
\label{2.4}
\end{eqnarray}
where we have defined the following parameters:
\begin{eqnarray}
\mu^{2}&=&m^{2}-\frac{\mathcal{E}^{2}}{1-\epsilon};\nonumber\\
\delta&=&2m\chi;\label{2.5}\\
\tau^{2}&=&\frac{l^{2}}{\eta^{2}}+\chi^{2}\nonumber.
\end{eqnarray}

Hence, by following the steps from Eq. (\ref{1.17}) to Eq. (\ref{1.20}), we obtain the following relativistic energy levels
\begin{eqnarray}
\mathcal{E}_{n,\,l}=\pm\sqrt{\left(1-\epsilon\right)\,m^{2}\left[1-\frac{4\chi^{2}}{\left[2n+2\left|\frac{l^{2}}{\eta^{2}}+\chi^{2}\right|+1\right]^{2}}\right]}.
\label{2.6}
\end{eqnarray}

Hence, Eq. (\ref{2.6}) are the allowed energy of a relativistic scalar particle confined to the scalar potential established in Eq. (\ref{1.11}) in the cosmic string spacetime subject to the effects of the Lorentz symmetry violation background defined by a fixed time-like 4-vector given in Eq. (\ref{2.1}). In this case, we have seen that the fixed time-like 4-vector yields a change in the component $g_{tt}$ of the metric tensor and, as a consequence, the Klein-Gordon equation in the cosmic string spacetime is modified. We can observe that the allowed energies (\ref{2.6}) are modified due to the effects of the Lorentz symmetry violation in contrast to that obtained in Eq. (\ref{1.20}). This change in the allowed energies stems from the Lorentz symmetry violation background to be defined by a fixed time-like 4-vector in the case of Eq. (\ref{2.6}) and by a fixed space-like 4-vector in the case of Eq. (\ref{1.20}). 

On the other hand, in both cases given by Eqs. (\ref{2.6}) and (\ref{1.20}), the effects of the topology of the cosmic string spacetime is given by the presence of the parameter $\eta$. Note that the allowed energies (\ref{2.6}) do not change by performing the change $\chi\rightarrow-\chi$. Moreover, by taking the limit as $\eta\rightarrow1$ in Eq. (\ref{2.6}), therefore, we also have the allowed energies of the relativistic scalar particle confined to the the scalar potential (\ref{1.11}) under Lorentz symmetry breaking effects in the Minkowski spacetime, but with a spectrum of energy that differs from that given in Eq. (\ref{1.20}). Finally, by taking $\epsilon=0$, the effects of the Lorentz symmetry violation vanish and we recover the results discussed in the previous section.

\section{conclusions}

We have investigated the behaviour of a relativistic scalar particle confined to a scalar potential subject to the effects of the Lorentz symmetry violation in the cosmic string spacetime. We have established two possible scenarios of the violation of the Lorentz symmetry in the cosmic string spacetime defined by  
a fixed space-like 4 vector and a fixed time-like 4-vector $\xi_{a}$ that satisfies the condition $\xi_{\mu}\left(x\right)\,\xi^{\mu}\left(x\right)=\mathrm{const}$, established in Refs. \cite{curv,curv2,curv3}. In both cases, we have shown that the Klein-Gordon equation in the cosmic string spacetime is modified due to the effects of the Lorentz symmetry violation background that change the components of the metric tensor. In addition, we have shown that bound state solutions to the Klein-Gordon equation in the cosmic string spacetime can be achieved in both cases, but the allowed energies differ from each other.

Furthermore, we have seen the influence of the topology of the cosmic string spacetime on the allowed energies obtained in this work. By taking the limit as $\eta\rightarrow1$ in the relativistic energy levels, therefore, we have obtained in both cases the allowed energies of the relativistic scalar particle confined to a scalar potential under Lorentz symmetry breaking effects in the Minkowski spacetime.

\acknowledgments{The authors would like to thank A. Y. Petrov, C. Furtado and J. A. Helay\"el-Neto for interesting discussions, and CNPq (Conselho Nacional de Desenvolvimento Cient\'ifico e Tecnol\'ogico - Brazil) for financial support.}

\end{document}